\documentclass[aps,english,twocolumn,superscriptaddress]{revtex4}
\usepackage{graphicx}
\usepackage{amssymb}
\usepackage{babel}

\begin{document}

\title{Effects of static charging and exfoliation of layered crystals}

\author{M. Topsakal}
\affiliation{UNAM-Institute of Materials Science and
Nanotechnology, Bilkent University, Ankara 06800, Turkey}
\author{S. Ciraci}\email{ciraci@fen.bilkent.edu.tr}
\affiliation{UNAM-Institute of Materials Science and
Nanotechnology, Bilkent University, Ankara 06800, Turkey}
\affiliation{Department of Physics, Bilkent University, Ankara
06800, Turkey}

\date{\today}

\begin{abstract}
Using first-principle plane wave method we investigate the effects of static
charging on structural, electronic and magnetic properties of suspended, single
layer graphene, graphane, fluorographene, BN and MoS$_2$ in honeycomb structure.
The limitations of periodic boundary conditions in the treatment of negatively charged layers are
clarified. Upon positive charging the band gaps between the conduction and valence
bands increase, but the single layer nanostructures become metallic owing to the Fermi
level dipping below the maximum of valence band. Moreover, their bond lengths increase
leading to phonon softening. As a result, the frequencies of Raman active modes are
lowered. High level of positive charging leads to structural instabilities in single
layer nanostructures, since their specific phonon modes attain imaginary frequencies.
Similarly, excess positive charge is accumulated at the outermost layers of metallized
BN and MoS$_2$ sheets comprising a few layers. Once the charging exceeds a threshold value the
outermost layers are exfoliated. Charge relocation and repulsive force generation are
in compliance with classical theories.

\end{abstract}

\maketitle

\section{introduction}
Single layer graphene,\cite{novo} graphane CH,\cite{graphan,hasanch}
fluorographene CF,\cite{nair,hasancf} BN\cite{BN,mehmetbn} and
MoS$_2$\cite{mos2,canmos2} have displayed unusual chemical and physical
properties for future nanotechnology applications. Furthermore, the properties
of these nanomaterials can be modified by creating excess electrical charge. For
example, linear crossing of bands of graphene at the Fermi level gives rise to
electron-hole symmetry, whereby under bias voltage the charge carriers can be
tuned continuously between electrons and holes in significant
concentrations.\cite{dirac} This way, the conductivity of graphene can be
monitored. Similar situation leading to excess electrons or holes can also be
achieved through doping with
foreign atoms.\cite{wang,wehling,sevincli,ataca,topsakal} Layered
materials can be exfoliated under excessive charging, which is created by
photoexcitation for very short time.\cite{carbone,miyamoto} It is proposed that
the femtosecond laser pulses rapidly generate hot electron gas, which spills out
leaving behind a positively charged graphite slab. Eventually, charged outermost
layers of graphite are exfoliated.\cite{miyamoto}

Recently, the effects of charging of graphene have been treated in different
studies. Ekiz \textit{et al.}\cite{ekiz} showed that oxidized graphene domains, which
become insulator upon oxidation, change back to the metallic state using
electrical stimulation. Theoretically, based on the first principles
calculations, it has been shown that the binding energy and magnetic moments of
adatoms adsorbed to graphene can be modified through static
charging.\cite{topsakal,cohen} Possibility of transforming the electronic structure
of one species to another through gating modeled by charging has been pointed
out.\cite{cohen2} It is argued that diffusion of adsorbed oxygen
atoms on graphene can be modified through charging.\cite{sofo} We found that
pseudopotential plane wave calculations of charged surfaces using periodically
repeating layers are sensitive to the vacuum spacing between adjacent cells and
have limited applicability.\cite{topsakal}

In this paper we investigate the effect of static charging on suspended
(or free standing) single layer nanostructures, such as graphene, graphane (CH),
fluorographene (fully fluorinated graphene) (CF), boron nitride (BN) and
molybdenum disulfide (MoS$_2$). All these honeycomb nanostructures have two dimensional
(2D) hexagonal lattice. First, we examine how the size of the "vacuum" potential
between layers affects the calculated properties of the \textit{negatively charged} single-layer
nanostructures when treated using periodic boundary conditions. We then investigate
the effect of charging on the electronic energy band structure and atomic structure.
We show that the bond lengths and hence 2D lattice constants increase as a result
of electron removal from the single layer. Consequently, phonons soften and the
frequencies of Raman active modes are lowered. As a result of electron removal,
three-layer, wide band gap BN and MoS$_2$ sheets are metallized and excess positive
charge is accumulated mainly at the outermost atomic layers. Owing to Coulomb
force those layers start to repel each other. When exceeded the weak van der Waals (vdW)
attraction, the repulsive Coulomb force initiates the exfoliation.

\section{Method}
The present results are obtained by performing first-principles plane wave
calculations carried out within spin-polarized and spin-unpolarized density
functional theory (DFT) using  projector-augmented wave potentials.\cite{paw}
The exchange correlation potential is approximated by Generalized Gradient
Approximation.\cite{PW91} For a better account of weak interlayer attraction in
layered crystals, van der Waals (vdW) interaction is also taken into
account.\cite{grimme} A plane-wave
basis set with kinetic energy cutoff of 500 eV is used. All atomic positions and
lattice constants are optimized by using the conjugate gradient method, where
the total energy and atomic forces are minimized. The convergence for energy is
chosen as 10$^{-5}$ eV between two steps, and the maximum force allowed on each
atom is less than 0.01 eV/\AA{}. The Brillouin zone (BZ) is sampled by (15x15x5)
special \textbf{k}-points for primitive unit cell. Calculations for neutral, as
well as charged systems are carried out by using VASP package.\cite{vasp}

Two-dimensional single layers or slabs and a vacuum space $s$ between them are
repeated periodically along the perpendicular $z$-direction. The amount of
charging is specified as either positive charging, i.e. electron depletion ($Q >0$),
or negative charging, i.e. excess electrons ($Q < 0$), in units of $\pm$
electron (e) per unit cell. Average surface charge density is specified as $\bar{\sigma}= Q/A$,
i.e the charge per unit area, $A$, being the area of the unitcell.
Normally, periodic boundary conditions realized by repeating charged supercells
has a divergent electric potential energy and has
drawbacks and limitations, which have been the subject matter of
several studies in the past. To achieve the convergence of electronic potential, additional
neutralizing background charge is applied.\cite{leslie,payne} Recently, error
bars in computations due to compensating charge have been estimated.\cite{cohen2}
The dipole corrections can be carried out for cubic structures, if a finite electric
dipole moment builds in the unit cell.\cite{dabo,schefler} Monopole and dipole 
corrections are also treated self-consistently.\cite{gava} Various charged 
structures have been also treated by using different approaches and computational
methods.\cite{fu,blochl,schutz,heinze,lozovoi,filhol,schnur} Owing to those
theoretical advances, studies on charged systems can now reveal useful information, 
when treated carefully.

\section{Charging of suspended single layers}

\begin{figure}
\includegraphics[width=8cm]{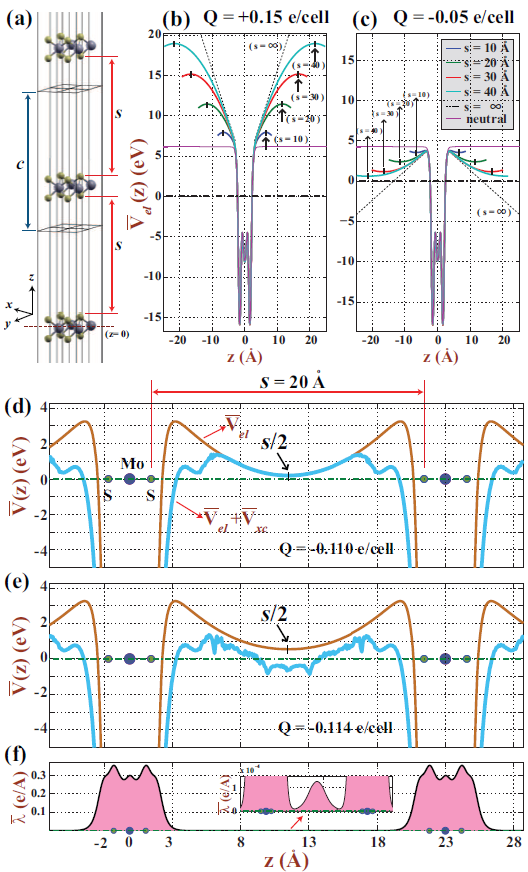}
\caption{(Color online) (a) Description of supercell geometry used to treat 2D
single layer MoS$_2$. $c$ and $s$ are supercell lattice constant and vacuum
spacing between adjacent layers. The $z$-axis is perpendicular to the layers. (b)
Self-consistent potential energy of positively charged ($Q>0$ per cell), periodically repeating
MoS$_{2}$ single layers, which is planarly averaged along $z$-direction.
$\bar{V}_{el}(z)$ is calculated using different vacuum spacings $\textit{s}$
as specified by inset. The planarly averaged potential energy of a single and infinite
MoS$_2$ layer is schematically shown by linear dashed lines in the vacuum region.
The zero of energy is set at the Fermi level indicated by dash-dotted lines.
(c) $\bar{V}_{el}(z)$ of negatively charged ($Q<0$ per cell) and periodically repeating
MoS$_2$ single layers. Averaged potential energy of infinite MoS$_2$ single layer is shown by linear
and dashed line in the vacuum region. (d) Variation of  $\bar{V}_{el}(z)$  and total
potential energy including electronic and exchange-correlation potential,
$\bar{V}_{el}(z)$ + $\bar{V}_{xc}(z)$, between two negatively charged MoS$_2$
layers corresponding to $Q$=-0.110 e/unitcell before the spilling of electrons into vacuum.
The spacing between MoS$_2$ layers is $s=20$ \AA. (e) Same as (d) but $Q$=-0.114
e/unitcell, where the total potential energy dips below $E_F$ and hence excess electrons
start to fill the states localized in the potential well between two MoS$_2$ layers.
(f) Corresponding planarly averaged charge density $\bar{\lambda}$. Accumulation of
the charge at the center of $s$ is resolved in a fine scale. Arrows indicate the
extremum points of  $\bar{V}_{el}(z)$ in the vacuum region for $Q>0$ and $Q<0$ cases.}

\label{figure-structure}
\end{figure}

The negative and positive charging of suspended single layer graphene, CH, CF,
BN and MoS$_2$ are treated using supercell method. In Fig.~\ref{figure-structure}
(a) we describe MoS$_2$ single layers, which are periodically repeated along
the $z$-direction and separated by a vacuum spacing $s$ between the adjacent
outermost sulfur planes. In Fig.~\ref{figure-structure} (b) and (c) the
self-consistent electronic potential energy, $V_{el}(\textbf{r})$ is averaged in the
planes perpendicular to the $z$-axis to obtain planarly averaged 1D potential energy
$\bar{V}_{el}(z)$ for different values of $s$.

In the vacuum region, the electronic potential energy $\bar{V}_{el}(z)$
strongly depends on the vacuum spacing $s$. For an infinitely large single
plane having excess charge $Q>0$ per cell, the potential energy in the vacuum
region is linear, if it is not periodically repeating. Thus, as $z \rightarrow \infty$,
$\bar{V}_{el}(z \rightarrow \infty) \rightarrow +\infty$ as schematically shown
in Fig.~\ref{figure-structure} (b). However, for a periodically repeating single
layers (within the periodic boundary conditions) the potential energy is symmetric
with respect to the center of vacuum spacing and it passes through a maximum at $s$/2.
The maximum value of the potential increases with increasing $s$ in
Fig.~\ref{figure-structure}(b).

In contrast, for a negatively ($Q<0$ per cell) charged and infinite MoS$_2$ single layer, a
reverse situation  occurs as shown in Fig.~\ref{figure-structure} (c). Namely $\bar{V}_{el}(z
\rightarrow \infty) \rightarrow -\infty$ linearly, if MoS$_2$ single layer is not periodically
repeating. Notably, the energy of a finite size, single layer nanostructure (i.e. a flake)
does not diverge, but has finite value for large $z$ both for $Q>0$ and $Q<0$ cases. On the other
hand, for periodically repeating single layers within the periodic boundary conditions,
potential energies are symmetric with respect to the center of vacuum spacing and they
passes through a minimum at $s/2$. This way a potential well is formed in the vacuum region
between two adjacent layers. Normally, the depth of this well increases with increasing 
negative charging and $s$. At a critical value of negative charge, the self-consistent
potential energy $V(\textbf{r})$ including electronic and exchange-correlation
potential energies dip below the Fermi level (even if $\bar{V}_{el}(z) > E_F$) and
eventually electrons start to occupy the states localized in the quantum well.
Such a situation is described in  Fig.~\ref{figure-structure} (d)-(f). Of course,
this situation is an \textit{artifact} of the method using plane wave basis set 
and the repeating layers separated by the vacuum space $s$. 
Despite that, the method may provide reasonable description of the
negatively charged layers until the minimum of the well dips below the Fermi level.
According to this picture, the escaping of electrons out of the material is delayed
in relatively short $s$. On the other hand, the interaction between layers prevents
one from using too short $s$. Earlier, this limitation of the method is usually overlooked.
The critical value of negative charge depends on $s$ value. It should be noted that 
for $s$=20 \AA, electrons start to escape from the graphene layer for $Q$=-4.03x10$^{13}$ e/cm$^{2}$,
even though larger doping of 4x10$^{14}$ e/cm$^{2}$ has been attained for garphene
on SiO$_{2}$ substrate.\cite{efetov}

\begin{figure}
\includegraphics[width=8.3cm]{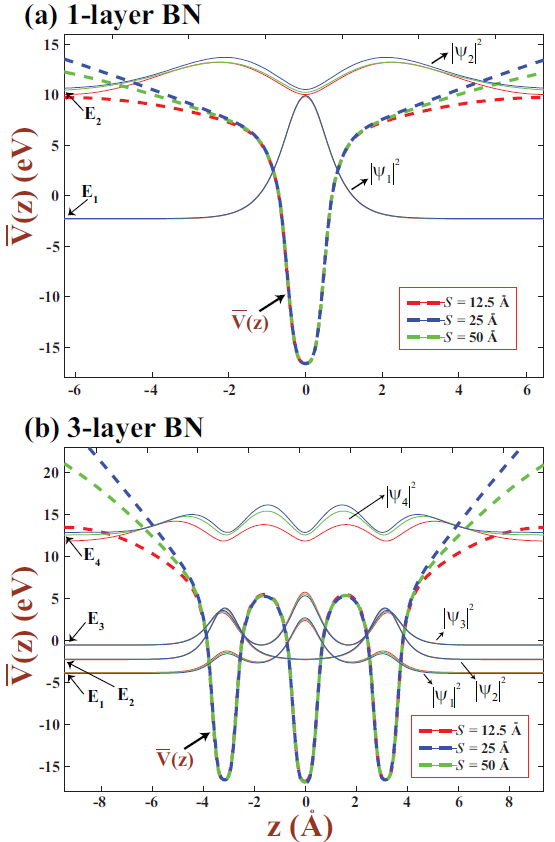}
\caption{(Color Online) (a) Energy eigenvalues of the occupied electronic states, $E_{i}$ and
corresponding $|\Psi_{i}(z)|^2$ are obtained by the numerical solution of the Schrodinger
equation of a planarly averaged, 1D electronic potential energy of single layer graphene
for $s$=12.5 $\AA$, 25 $\AA$ and 50 $\AA$ shown by dashed lines. (b) Same as (a) for 3-layer graphene. Zeros of
$|\Psi_{i}(z)|^2$ at large $z$ are aligned with the corresponding energy eigenvalues.}
\label{figure-schrodinger}
\end{figure}

In the case of positive charging, even if $\bar{V}_{el}(z)$ is not linear and does
not increase to $+\infty$, the periodic boundary conditions using sufficiently large
$s$ can provide a realistic description of charged systems, since the
wave functions in the vacuum region rapidly decay under high and wide potential barrier.
Therefore, the calculated wave functions and electronic energies are not affected
even if $\bar{V}_{el}(z)$ is smaller than the electronic potential corresponding
to infinite vacuum spacing. We demonstrate our point of view by solving
directly the Schrodinger equation to obtain the wave functions and energy eigenvalues
for the planarly averaged 1D potentials of single layer and 3-layer graphene corresponding
to $s$=12.5, 25, 50 $\AA$ in Fig.~\ref{figure-schrodinger}. One sees that the large
difference, $\Delta \bar{V}_{el}(z)=\bar{V}_{el,s=50 \AA}(z)-\bar{V}_{el,s=12.5 \AA}(z)$ do not
affect the occupied states at their tail region in the vacuum spacing; the energy
difference is only 5 meV (which cannot be resolved from the figure) between smallest
and largest vacuum spacing $s$, which is smaller than the accuracy limit of DFT calculations.
As one expects, the dependence on the vacuum spacing increases for excited states, which have
relatively larger extension and hence they are affected from $\Delta \bar{V}_{el}(z)$.

\begin{figure}
\includegraphics[width=8.3cm]{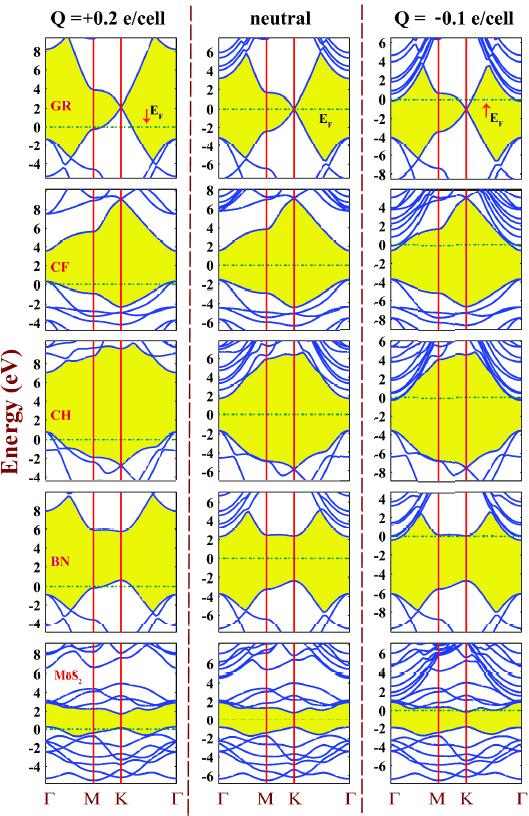}
\caption{(Color Online) Energy band structures of 2D single layer of graphene C,
fluorographene CF, graphane CH, BN and MoS$_2$ calculated for $Q =+0.2$ e/cell,
$Q = 0$ (neutral) and $Q = -0.10$ e/cell. Zero of energy is set at the Fermi
level indicated by dash-dotted lines. The band gap is shaded. Note that band gap
increases under positive charging. Parabolic bands descending and touching the Fermi
level for $Q<0$ are free electron like bands. Band calculations are carried out for
$s$=20 \AA.}

\label{figure-bands}
\end{figure}

By taking the above limitations of the method in negative charging into account,
we now examine the effect of charging of single layers of graphene, CF, CH,
BN and MoS$_2$ on their electronic structure and bond lengths. In
Fig.~\ref{figure-bands} the changes in band structure with charging are
significant within DFT. For example, the band gap (i.e. the energy gap between
the top of the valence band and the minimum of the conduction band) of neutral
single layer BN  increases from 4.61 eV to 5.12  and to 5.54 eV as $Q$ increases
from $Q$=0 to +0.2 e/cell and to +0.4 e/cell, respectively. The increase of the band gap
occurs due to the fact that the electronic potential energy becomes deeper with
increasing electron depletion. For $Q > 0$, the Fermi level dips in the valance
band and creates holes.

In contrast, parabolic free electron like bands, which occur above the vacuum level in
the neutral case, start to descend as a result of negative charging ($Q<0$) and eventually
they touch the Fermi level. Upon increasing $Q$ these parabolic bands
start to be occupied and hence part of $Q$ is transferred to the quantum well
in the vacuum region. This way the rate of accumulation of excess charge in the conduction band
of single layer nanostructure recedes. Even if these parabolic bands appear as touching the Fermi
level in the same band structure in ($k_{x},k_{y}$)-plane they are physically separated from
the states of single layer honeycomb structure under study. As it was mentioned before, this
situation is an artifact of periodic boundary conditions and can be viewed as the
vanishing of the work function. We note that in the case of negatively charged, finite size single-layer, 
the excess electrons are hindered from spilling out to the vacuum by a wide tunneling barrier, 
even if $\bar{V}_{el}(z)$ is lowered below the Fermi level in vacuum for large $z$.

Incidentally, for both $Q>0$ and $Q<0$, the spin-polarized calculations carried out for
single layers of graphene, CH, CF, BN and MoS$_2$ did not yield any magnetic state
as a result of charging.

\begin{figure}
\includegraphics[width=8cm]{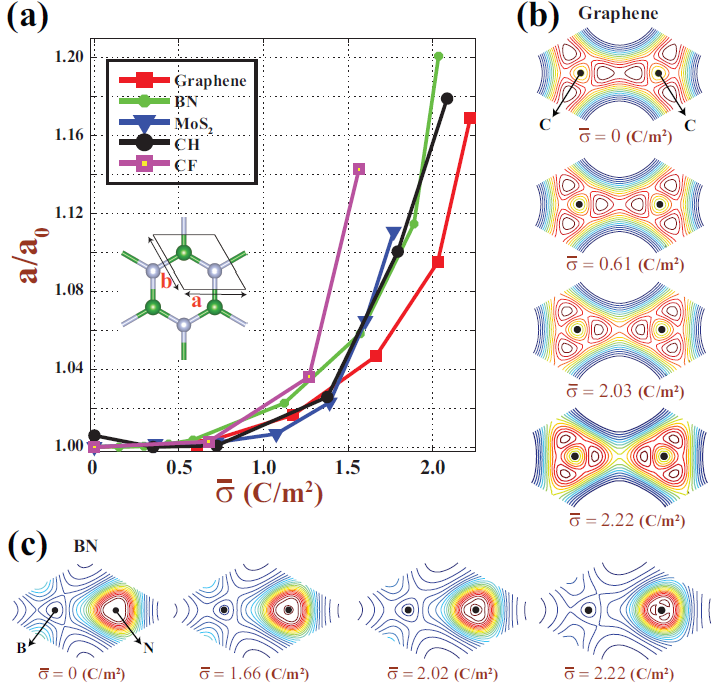}
\caption{(Color Online) (a) Variation of the ratio of lattice constants $a$ of
positively charged single layer graphene, BN, CH, CF and MoS$_2$  to their
neutral values $a_o$ with the average surface charge density,  $\bar{\sigma}$.
The unit cell and the lattice vectors are described by inset. (b) The charge
density contour plots in a plane passing through a C-C bond. (c) Same as B-N
bond.}

\label{figure-lattice}
\end{figure}

Another crucial effect of charging appears in the variation of the bond lengths
with $Q$. As shown in Fig.~\ref{figure-lattice} (a) the bond length or
lattice constants of single layer graphene, BN, CH, CF and MoS$_2$ increase with
increasing positive charge density $\bar{\sigma}$. The elongation of the
bond length is slow for small $\bar{\sigma}$, but increases quickly when
$\bar{\sigma} \gtrsim 1$ C/m$^2$. The bonds get longer when the electrons are
removed from the system and hence bonds become weaker. The contour plots of
total charge density in a plane passing through C-C and B-N bonds of graphene
and BN honeycomb structures in Fig.~\ref{figure-lattice} (b) and (c),
show that the charge density between atoms becomes weaker with increasing electron
depletion. Weakening of bonds can have crucial consequences as phonon softening
and is observable through Raman spectrum. In fact, the Raman active mode of
graphene calculated by using density functional perturbation theory is at
1524 cm$^{-1}$ and shifts down to 1510 cm$^{-1}$ for $Q$=+0.2 e/cell, and to 1311
cm$^{-1}$ for $Q$=0.4 e/cell. To confirm whether the elongation of bonds
dominates the Raman shift of graphene, we calculated the Raman active modes of
neutral graphene having the same lattice constant of graphene when charged by
$Q$=+0.4 e/cell. In this case the Raman active mode shifted to 1274 cm$^{-1}$,
which is close to the Raman active mode calculated for graphene charged with
$Q$=+0.4 e/cell. We also note that the excessive charging of single layer
materials considered in this paper lead to instability. This is revealed
from phonon dispersion calculations. For example, neutral graphene which has
normally stable planar structure and positive frequencies of acoustical branches
in BZ, starts to attain imaginary frequencies of long wavelength acoustical
modes at excessive charging. Weakening of graphene layer is expected to be reflected
to its elastic properties, in particular to its stiffness.\cite{stiffness}

\section{Exfoliation of layered BN and MoS$_2$}

\begin{table}
\caption{Dependence of the threshold charges on the vacuum spacing $s$ (\AA) between 3-layer slabs.
Threshold charge, $Q_e$ (e/cell) where exfoliation sets in and corresponding threshold
average surface charge density $\bar{\sigma}_e=Q_{e}/A$ (C/m$^2$) are calculated
for positive charged 3-layer Graphene, BN and MoS$_2$ sheets for $s$=50 \AA~ and $s$=20 \AA. The
numbers of valence electrons per unit cell of the slab are also given in the second column.}
\label{table3} \centering{}
\begin{tabular}{|c|c|c|c|}
\hline
System & \# of e & $Q_{e}$ (e/cell) & $\bar{\sigma}$ (C/m$^{2}$)
\tabularnewline
\hline
\hline
3-layer Graphene (s=50) & 24 & +0.160  & +0.49  \tabularnewline
\hline
3-layer Graphene (s=20) & 24 & +0.205  & +0.62  \tabularnewline
\hline
3-layer BN (s=50) & 24 & +0.225  & +0.66  \tabularnewline
\hline
3-layer BN (s=20) & 24 & +0.320  & +0.94  \tabularnewline
\hline
3-layer MoS$_2$ (s=50) & 54 & +0.322  & +0.57  \tabularnewline
\hline
3-layer MoS$_2$ (s=20) & 54 & +0.480  & +0.86  \tabularnewline
\hline

\end{tabular}
\end{table}

We next investigate the exfoliation of single layer BN and MoS$_2$ from their
layered bulk crystal through charging. We model 3-layer slab (sheet) of BN and MoS$_2$
as part of their layered bulk crystal. We considered only 3-layer slabs in order to cut
the computation time, since the model works also for thicker slabs consisting
of 6-10 layers graphene.\cite{topsakal}  Energy minimizations of neutral sheets relative to stacking
geometry are achieved. Stacking of 3-layer BN and MoS$_2$ slabs comply with the stacking
of layers in 3D layered BN\cite{mehmetbn} and MoS$_2$ crystals.\cite{canmos2}
In these slabs, the layers are hold together mainly by attractive vdW
interactions of a few hundreds meV and any repulsive interaction overcoming it
leads to exfoliation. When electrons are injected to or removed from the slab, the
Fermi level shifts up or down and cross the conduction or valance band of the
insulator and attribute to it a metallic character. At the end, the excess
charge by itself accumulates on the surfaces of the metallic slab inducing a
repulsive Coulomb interaction between the outermost layers of the slab. Here we consider
positive charging only, since in the case of negative charging the excess charges
quickly spill into the vacuum before the exfoliation sets in.

\begin{figure}
\includegraphics[width=7cm]{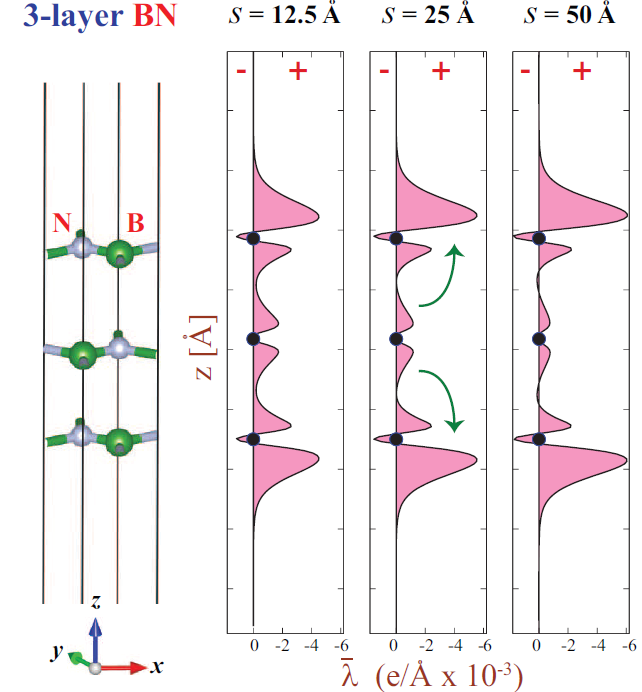}
\caption{(Color Online) Variation of planarly averaged positive excess charge $\bar{\lambda}(z)$ along
$z$-axis perpendicular to the BN-layers calculated for different $s$. As $s$ increases more excess
charge is transferred from center region  to the surface planes.}
\label{figure-bn-charge}
\end{figure}

\begin{figure}
\includegraphics[width=9cm]{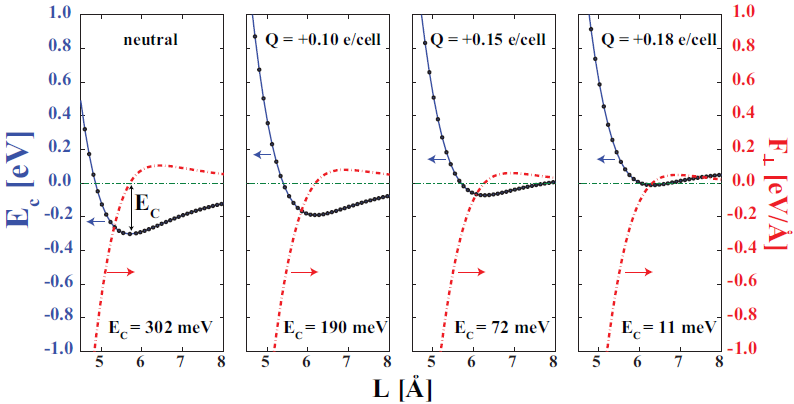}
\caption{(Color Online) Variation of cohesive energy and perpendicular force
$F_{\perp}$ in 3-layer BN slab as a function of the distance $L$ between the
surfaces. Energies and forces are calculated for different levels of excess
positive charge $Q$ (e/unit cell). Zero of energy corresponds to the energy as
$L \rightarrow \infty$. }

\label{figure-energy}
\end{figure}

The amount of charge in the unit cell, which is necessary for the onset of
exfoliation, is defined as the threshold charge $Q_e$.
Threshold charges are calculated for 3-layer slabs of graphene, BN and MoS$_2$
for $s$=20 \AA~ and $s$=50 \AA. Results presented in Tab.\ref{table3} indicate
that the amount of threshold charge decreases with increasing $s$. This confirms
our arguments in Sec. III that in positive charging large vacuum space, $s$, is
favored. The mechanism underlying this finding is summarized in Fig.~\ref{figure-bn-charge}
where we show the linear charge density, $\bar{\lambda}(z)$ calculated for
different $s$ values of a 3-layer BN. For small $s$, the excess charge
accumulates mainly at surfaces of the slab, also with some significant fraction
inside the slab. However, as $s$ increases some part of $Q$ is transferred from
inside to the outer surface giving rise to the increase of the charge accumulation
at the surface. At the end, for the
same level of charging the induced Coulomb repulsion increases with increasing
$s$. Accordingly, the same slab requires relatively smaller amount of threshold
charge $Q_e$ to be exfoliated, if $s$ is taken large.

\begin{figure*}
\includegraphics[width=15cm]{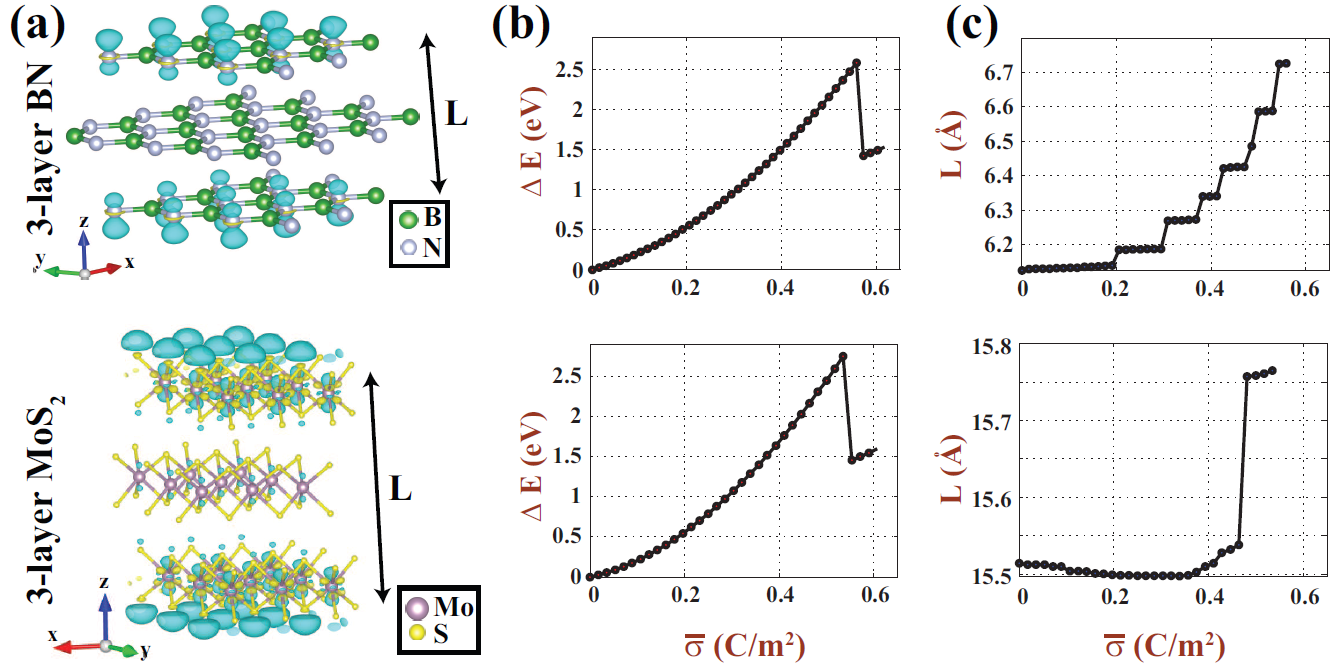}
\caption{(Color Online) Exfoliation of outermost layers from layered BN and
MoS$_2$ slabs by positively charging of slabs. (a) Turquoise isosurfaces of excess
positive charge density. (b) Change in total energy with excess
surface charge density. (c) Variation of $L$ of slabs with charging. }

\label{figure-exfoliation}
\end{figure*}

In Fig.~\ref{figure-energy} we present the variation of the cohesive energy of the
3-layer BN slab relative to three free BN layers for neutral $Q=0$ and positive
charged $Q >0$ cases as a function of the distance $L$ between the outermost BN atomic
planes of 3-layer BN slab. The cohesive energy for a given $L$ is obtained from
the following expression: $E_{C}= E_{T}$ [3-Layer BN] - 3$E_{T}$ [single layer BN].
The total energy of the single layer BN, $E_{T}$ [single layer] is calculated in a smaller
supercell to keep the density of the background charge the same. The cohesive
energy of the neutral slab in equilibrium is $\sim$ 302 meV/cell. If the spacings of
layers (i.e. $L$) starts to increase, an attractive force $F_{\perp}=-\partial
E_{T}/\partial L$ acts to restore the equilibrium spacing. $F_{\perp}(L)$ first
increases with increasing $L$, passes through a maximum and then decays to zero.
In Fig.~\ref{figure-energy} we also
show how the minimum of cohesive energy decreases and moves to relatively large
spacings with increasing $Q$. Concomitantly, the maximum of the attractive force
for a given $Q$, $F_{\perp,max}$ decreases with increasing $Q$ and eventually
becomes zero. This give rise to the exfoliation. We note that despite the limitations
set by the neutralizing uniform charge on the total energy, the cohesive energies
calculated for different charge levels reveal useful qualitative information on
the effects of charging.

In Fig.~\ref{figure-exfoliation} (a) we show
isosurfaces of excess positive charge densities of 3-layer BN and MoS$_2$ slabs.
These slabs become metallic upon extracting electrons (i.e. upon positive charging) and
excess charges reside at both surfaces of slabs. As shown in Fig.~\ref{figure-exfoliation}
(b), the total energy raises with increasing charging or average charge
density, $\bar{\sigma}$. In compliance with Fig.~\ref{figure-energy}, the separation
between surface layers, $L$, increases. The sharp drop of $\Delta E$ at $Q_e$ or
$\bar{\sigma}_e$ indicate the onset
of exfoliation due to the repulsive Coulomb force pulling them to exfoliate.
In Fig.~\ref{figure-exfoliation} (c) $L$ increases with increasing  charging as discussed in
Fig.~\ref{figure-energy}. The increments of $L$ exhibits a stepwise behavior
for BN. This is also artifact of the method, where forces are calculated within preset threshold values.

The variation of $L$ of MoS$_2$ slab with $Q > 0$  display a different
behavior due to charge transfer from Mo to S atoms. The exfoliation
due to the static charging can be explained by a simple electrostatic model, where
the outermost layers of slabs is modeled by uniformly charged planes, which yields
repulsive interaction independent of their separation distance, i.e. $F \propto Q^{2}/(A\cdotp
\epsilon_{0})$, where $\epsilon_{0}$ is static dielectric constant.\cite{topsakal} Calculated
forces differ from the classical force due to screening effects of excess charge residing inside
the slabs.

\section{Discussions and conclusions}

In this study, the threshold values of static charge, $Q_e$, to be implemented in
the slabs to achieve exfoliation are quite high. Such high static charging of
layers can be achieved locally through the tip of Scanning Tunnelling Microscope
or electrolytic gate.\cite{efetov}
The dissipation of locally created excess charge in materials may involve a
decay time $\tau_D$. Relatively longer $\tau_D$ can induce a local instability and
the desorption of atoms from nanoparticles. Experimentally ultra-fast graphene ablation
was directly observed by means of electron crystallography.\cite{carbone}
Carriers excited by ultra-short laser pulse transfer energy to strongly coupled
optical phonons. Graphite undergoes a contraction, which is subsequently
followed by an expansion leading eventually to laser-driven
ablation.\cite{carbone} Much recently, the understanding of photoexfoliation have
been proposed, where exposure to femtosecond laser pulses has led to athermal
exfoliation of intact graphenes.\cite{miyamoto} Based on time dependent DFT
calculations (TD-DFT), it is proposed that the femtosecond laser pulse rapidly
generates hot electron gas at $\sim20.000$ K, while graphene layers are
vibrationally cold. The hot electrons spill out, leaving behind a positively
charged graphite slab. The charge deficiency accumulated at the top and bottom
surfaces lead to athermal excitation.\cite{miyamoto} The exfoliation in static
charging described in Fig.~\ref{figure-exfoliation} is in compliance with the
understanding of photoexcitation revealed from previous TD-DFT calculations,\cite{miyamoto}
since the driving force which leads to the separation of graphenes from graphite is mainly
related with electrostatic effects in both methods.

In summary, the present study investigated the effects of charging on the structural
and electronic properties of single layer graphene, graphene derivatives, BN and
MoS$_2$, which have honeycomb structure. We concluded that while caution has to be
exercised in the studies involving negative charging using large vacuum spacing, positive
charging can be treated safely using large vacuum spacing.

We found that upon positive charging the band gaps of single layers of BN
and MoS$_2$ increase and the unit cells are enlarged. Consequently the phonons
become softer. The charging of BN and MoS$_2$ slabs were also studied. While
these slabs are wide band semiconductors, they become metallic upon positive
charging. Consequently, excess charges are accumulated on the surfaces of slabs
and induce repulsive force between outermost layers. With increasing positive
charging the spacing between these layers increases, which eventually ends with
exfoliation.

\begin{acknowledgments}
We thank S. Cahangirov for helpful discussions. We acknowledge partial financial support from The Academy of Science of Turkey (TUBA) and TUBITAK through Grant No: 108234. Part of the computational resources has been provided by TUBITAK ULAKBIM, High Performance and Grid Computing Center (TR-Grid e-Infrastructure).

\end{acknowledgments}

\end{document}